\documentclass[a4paper,11pt]{article}
\usepackage{a4wide}

\usepackage{amsmath}
\usepackage{amsthm}
\newtheorem{theorem}{Theorem}
\newtheorem{lemma}{Lemma}

\usepackage[utf8]{inputenc}

\usepackage[T1]{fontenc}
\usepackage{color}

\usepackage{graphicx}
\usepackage{tikz}
\usetikzlibrary{arrows}
\usetikzlibrary{decorations.pathmorphing}
\usetikzlibrary{shapes.geometric}
\usetikzlibrary{calc}
\usetikzlibrary{matrix}
\usetikzlibrary{positioning}
\usetikzlibrary{decorations.pathreplacing}
\pgfmathsetmacro{\constantx}{1.2}
\pgfmathsetmacro{\constanty}{1.2}
\pgfmathsetmacro{\nodescale}{0.60}
\pgfmathsetmacro{\letterfigsc}{0.75}
\pgfmathsetmacro{\letterscale}{1}

\usepackage[colorinlistoftodos]{todonotes}

\DeclareMathOperator{\LCP}{LCP}

\newcommand{\Count}{\ensuremath\textsf{Count}}
\newcommand{\Predecessor}{\ensuremath\textsf{Predecessor}}
\newcommand{\Locate}{\ensuremath\textsf{Locate}}
\newcommand{\occ}{\ensuremath\mathrm{occ}}

\title{Deterministic Indexing for Packed Strings}

\author{Philip Bille\thanks{Supported by the Danish Research Council (DFF – 4005-00267, DFF – 1323-00178) and the Advanced Technology Foundation} \\\texttt{phbi@dtu.dk} \and Inge Li G\o rtz$^*$ \\\texttt{inge@dtu.dk} \and Frederik Rye Skjoldjensen\thanks{Supported by the Danish Research Council (DFF – 1323-00178)} \\\texttt{fskj@dtu.dk}}
\date{\today}

\begin{document}
\tikzset{ 
nodesrep/.style={
    rectangle,
    align=center,
    draw=black,
},
table1withindices/.style={
    matrix of nodes,
    row sep=30pt,
    column sep=-\pgflinewidth,
    font=\footnotesize,
    text depth=1pt,
    text height=5pt,
    text width=9pt,
    text centered,
    nodes={
        rectangle,
        draw=black,
        anchor=north,
    },
    nodes in empty cells,
    row 1/.style={
        nodes={
          draw=black,
          font=\tiny,
          text width=25pt
        }
    },
    row 2/.style={
        nodes={
          draw=black,
          font=\tiny,
          text width=25pt
        }
    },
    row 3/.style={
        nodes={
          draw=black,
          font=\tiny,
          text width=25pt
        }
    },
    row 4/.style={
      nodes={
        draw=white
      }
    }
},
table2/.style={
    matrix of nodes,
    row sep=15pt,
    column sep=-\pgflinewidth,
    font=\footnotesize,
    text depth=1pt,
    text height=5pt,
    text width=9pt,
    text centered,
    nodes={
        rectangle,
        draw=black,
        anchor=north,
    },
    nodes in empty cells,
    row 1/.style={
        nodes={
          draw=black,
          font=\tiny,
          text width=25pt
        }
    },
    row 2/.style={
        nodes={
          draw=black,
          font=\tiny,
          text width=25pt
        }
    },
    row 3/.style={
        nodes={
          draw=black,
          font=\tiny,
          text width=25pt
        }
    },
    row 4/.style={
      nodes={
        draw=white
      }
    }
}
}
\newcommand{\LA}[0]{\mathsf{SA}_S}
\newcommand{\T}[0]{\mathsf{T}_S}
\newcommand{\HT}[0]{\mathsf{HT}_S}
\newcommand{\comment}[1]{\todo{\footnotesize #1}}
\maketitle
\begin{abstract}
Given a string $S$ of length $n$, the classic string indexing problem is to preprocess $S$ into a compact data structure that supports efficient subsequent pattern queries. In the \emph{deterministic} variant the goal is to solve the string indexing problem without any randomization (at preprocessing time or query time). In the \emph{packed} variant the strings are stored with several character in a single word, giving us the opportunity to read multiple characters simultaneously. Our main result is a new string index in the deterministic \emph{and} packed setting. Given a packed string $S$ of length $n$ over an alphabet $\sigma$, we show how to preprocess $S$ in $O(n)$ (deterministic) time and space $O(n)$ such that given a packed pattern string of length $m$ we can support queries in (deterministic) time
$ 
O\left(m/\alpha + \log m + \log \log \sigma\right),
$
where $\alpha =  w / \log \sigma$ is the number of characters packed in a word of size $w = \Theta(\log n)$. Our query time is always at least as good as the previous best known bounds and whenever several characters are packed in a word, i.e., $\log \sigma \ll w$, the query times are faster.
\end{abstract}

\section{Introduction}
Let $S$ be a string of length $n$ over an alphabet of size $\sigma$. The \emph{string indexing problem} is to preprocess $S$ into a compact data structure that supports efficient subsequent pattern queries. Typical queries include \emph{existential queries} (decide if the pattern occurs in $S$), \emph{reporting queries} (return all positions where the pattern occurs), and \emph{counting queries} (returning the number of occurrences of the pattern).

The string indexing problem is a classic well-studied problem in combinatorial pattern matching and the standard textbook solutions are the suffix tree and the suffix array (see e.g., \cite{Gusfield1997,manber1993suffix,McCreight1976,Weiner1973}). A straightforward  implementation of suffix trees leads to an $O(n)$ preprocessing time and space solution that given a pattern of length $m$ supports existential and counting queries in time $O(m\log \sigma)$ and reporting queries in time $O(m\log \sigma + \occ)$, where $\occ$ is the number of occurrences of the pattern. The suffix array implemented with additional arrays storing longest common prefixes leads to a solution that also uses $O(n)$ preprocessing time and space while supporting existential and counting queries in time $O(m + \log n)$ and reporting queries in time $O(m + \log n + \occ)$. If we instead combine suffix trees with perfect hashing~\cite{FKS1984} we obtain $O(n)$ \emph{expected} preprocessing time and $O(n)$ space, while supporting existential and counting queries in time $O(m)$ and reporting queries in time $O(m + \occ)$. The above bounds hold assuming that the alphabet size $\sigma$ is polynomial in $n$. If this is not the case, additional time for sorting the alphabet is required~\cite{FCFM2000}. For simplicity, we adopt this convention in all of the bounds throughout the paper.

In the \emph{deterministic} variant the goal is to solve the string indexing problem without any randomization. In particular, we cannot combine suffix trees with perfect hashing to obtain $O(m)$ or $O(m + \occ)$ query times. In this setting Cole et al.~\cite{cole2006suffix} showed how to combine suffix trees and suffix array into the \emph{suffix tray} that uses $O(n)$ preprocessing time and space and supports existential and counting queries in $O(m + \log \sigma)$ time and reporting queries in $O(m + \log \sigma + \occ)$ time. Recently, the query times were improved by Fischer and Gawrychowski~\cite{fischer2015alphabet} to $O(m + \log \log \sigma)$ and $O(m + \log \log \sigma + \occ)$, respectively. 

In the \emph{packed} variant the strings are given in a \emph{packed representation}, with several characters in a single word~\cite{Bille2011, BBBGGW2014, Belazzougui2012, arimura2016packed}. For instance, DNA-sequences have an alphabet of size 4 and are therefore typically stored using 2 bits per character with 32 characters in a 64-bit word. On packed strings we can read multiple characters in constant time and hence potentially do better than the immediate $\Omega(m)$ or $\Omega(m + \occ)$ lower bound for existential/counting queries and reporting queries, respectively. In this setting Takagi et al.~\cite{arimura2016packed} recently introduced the \emph{packed compact trie} that stores packed strings succinctly and also supports dynamic insertion and deletions of strings. In a static and deterministic setting their data structure implies a linear space and superlinear time preprocessing solution that uses $O(\frac{m}{\alpha} \log \log n)$ and $O(\frac{m}{\alpha}\log \log n + \occ)$ query time, respectively.

In this paper, we consider the string indexing problem in the deterministic and packed setting simultaneously, and present a solution that improves all of the above bounds. 

\subsection{Setup and result}
We assume a standard unit-cost word RAM with word length $w = \Theta(\log n)$, and a standard instruction set including arithmetic operations, bitwise boolean operations, and shifts. All strings in this paper are over an alphabet $\Sigma$ of size $\sigma$. The \emph{packed representation} of a string $A$ is obtained by storing $\alpha =  w / \log \sigma$ characters per word thus representing $A$ in $O(|A|\log \sigma/w)$ words. If $A$ is given in the packed representation we simply say that $A$ is a \emph{packed string}. 

Throughout the paper let $S$ be a string of length $n$. Our goal is to preprocess $S$ into a compact data structure that given a packed pattern string $P$ supports the following queries. 
\begin{itemize}
\item[] $\Count(P)$: Return the number of occurrence of $P$ in $S$.
\item[] $\Locate(P)$: Report all occurrences of $P$ in  $S$.

\item[] $\Predecessor(P)$: Returns the predecessor of $P$ in $S$, i.e., the lexicographic largest suffix in $S$ that is smaller than $P$.
\end{itemize}
We show the following main result. 

\begin{theorem}\label{theo:main}
Let $S$ be a string of length $n$ over an alphabet of size $\sigma$ and let $\alpha =  w /  \log \sigma$ be the number of characters packed in a word.   Given $S$ we can build an index in $O(n)$ deterministic time and space such that given a packed pattern string of length $m$ we can support $\Count$ and $\Predecessor$ in time $O(\frac{m}{\alpha} + \log m + \log \log \sigma)$ and $\Locate$ in time $O(\frac{m}{\alpha} + \log m + \log \log \sigma + \occ)$ time.
\end{theorem}
Compared to the result of Fischer and Gawrychowski~\cite{fischer2015alphabet}, Thm~\ref{theo:main} is always at least as good and whenever several characters are packed in a word, i.e., $\log \sigma \ll w$, the query times are faster. Compared to the result of Takagi et al.~\cite{arimura2016packed}, our query time is a factor $\log \log n$ faster.

Technically, our results are obtained by a novel combination of previous techniques. Our general tree decomposition closely follows Fischer and Gawrychowski~\cite{fischer2015alphabet}, but different ideas are needed to handle packed strings efficiently. We also show how to extend the classic suffix array search algorithm to handle packed strings efficiently.

\section{Preliminaries}
\paragraph{Deterministic hashing and predecessor}
We use the following results on deterministic hashing and predecessor data structures.

\begin{lemma}[Ru{\v{z}}i{\'c} {\cite[Theorem 3]{ruvzic2008constructing}}]\label{lem:hash}
A static linear space dictionary on a set of $k$ keys can be deterministically constructed in time $O(k (\log \log k)^2)$, so that lookups to the dictionary take time $O(1)$.
\end{lemma}
Fischer and Gawrychowski \cite{fischer2015alphabet} use the same result for hashing characters. In our context we will apply it for hashing words of packed characters. 

\begin{lemma}[Fischer and Gawrychowski {\cite[Proposition 7]{fischer2015alphabet}}]\label{lem:pred}
A static linear space predecessor data structure on a set of $k$ keys from a universe of size $u$ can be constructed deterministically in $O(k)$ time and $O(k)$ space such that predecessor queries can be answered deterministically in time $O(\log \log u)$.
\end{lemma}

\paragraph{Suffix tree}
The suffix tree $\T$ of $S$ is the compacted trie over the $n$ suffixes from the string $S$. We assume that the special character $\$\not \in \Sigma$ is appended to every suffix of $S$ such that each string is ending in a leaf of the tree. The edges are sorted lexicographic from left to right. We say that a leaf \emph{represents} the suffix that is spelled out by concatenating the labels of the edges on the path from the root to the leaf. In the same way an internal node represents a string that is a prefix of at least one of the suffixes. For a node $v$ in $\T$, we say that the \emph{subtree} of $v$ is the tree induced by $v$ and all proper descendants of $v$. We distinguish between implicit and explicit nodes: implicit nodes are conceptual and refer to the original non branching nodes from the trie without compacted paths. Explicit nodes are the branching nodes in the original trie. When we refer to nodes that are not specified as either explicit or implicit, then we are always referring to explicit nodes. The lexicographic ordering of the suffixes represented by the leafs corresponds to the ordering of the leafs from left to right in the compacted trie. For navigating from node to child, each node has a predecessor data structure over the first characters of every edge going to a child. With the predecessor data structure from Lemma \ref{lem:pred} navigation from node to child takes $O(\log \log \sigma)$ time and both the space and the construction time of the predecessor data structure is linear in the number of children.

\paragraph{Suffix array}\label{sec:la}
Let $S_1,S_2,\ldots, S_n$ be the $n$ suffixes of $S$ from left to right. The suffix array $\LA$ of $S$ gives the lexicographic ordering of the suffixes such that $S_{\LA[i]}$ refers to the $i$th lexicographic greatest suffix of $S$. This means that for every $1< i \leq n$ we have that $S_{\LA[i-1]}$ is lexicographic smaller than $S_{\LA[i]}$. For simplicity we let $\LA[i]$ refer to the suffix $S_{\LA[i]}$ and we say that $\LA[i]$ represents the suffix $S_{\LA[i]}$.
Every suffix from $S$ with pattern $P$ as a prefix will be located in a consecutive range of $\LA$. This range corresponds to the range of consecutive leafs in the subtree spanned by the explicit or implicit node that represents $P$ in $\T$. We can find the range of $\LA$ where $P$ prefix every suffix by performing binary search twice over $\LA$. A naïve binary search takes $O(m\log n)$ time: We maintain the boundaries, $L$ and $R$, of the current search interval and in each iteration we compare the median string from the range $L$ to $R$ in $\LA$, with $P$, and update $L$ and $R$ accordingly. This can be improved to $O(m + \log n)$ time if we have access to additional arrays storing the value of the longest common prefixes between a selection of strings from $\LA$. We construct the suffix array from the suffix tree in $O(n)$ time.

\section{Deterministic index for packed strings}
In this section we describe how to construct and query our deterministic index for packed strings. This structure is the basis for our result in Thm~\ref{theo:main}. For short patterns where $m < \log_\sigma(n)-1$ we store tabulated data that enables us to answer queries fast. We construct the tables in $O(n)$ time and space and answer queries in $O(\log \log \sigma + \occ)$ time. For long patterns where $m \geq \log_\sigma(n)-1$ we use a combination of a suffix tree and a suffix array that we construct in $O(n)$ time and space such that queries take $O(m/\alpha + \log\log n + \occ)$ time. For $m \geq \log_\sigma(n)-1$ we have that $\log \log n = \log (\frac{\log n}{\log \sigma}\log\sigma) = \log\log_\sigma n + \log \log \sigma \leq \log (\log_\sigma n - 1) + 1 + \log \log \sigma \leq \log m + 1 + \log \log \sigma$. This gives us a query time of $O(m/\alpha + \log m +\log\log \sigma + \occ)$ for the deterministic packed index.
We need the following connections between $\T$ and $\LA$: For each explicit node $t$ in $\T$ we store a reference to the range of $\LA$ that corresponds to the leafs spanned by the subtree of $t$ and for each index in $\LA$ we store a reference to the corresponding leaf in $\T$ that represents the same string.

We first describe our word accelerated algorithm for matching patterns in $\LA$ that we need for answering queries on long patterns. Then we describe how to build and use the data structures for answering queries on short and long patterns.
\subsection{Packed matching in $\LA$}
We now show how to word accelerate the suffix array matching algorithm by Manber and Myers~\cite{manber1993suffix}. They spend $O(m)$ time reading $P$ but by reading $\alpha$ characters in constant time we can reduce this to $O(m/\alpha)$.  We let $\LCP(i,j)$ denote the length of the longest common prefix between the suffixes $\LA[i]$ and $\LA[j]$ and obtain the result in Lemma \ref{lem:la}.

\begin{lemma}\label{lem:la}
Given the suffix array $\LA$ over the packed string $S$ and a data structure for answering the relevant $\LCP$ queries, we can find the lexicographic predecessor of a packed pattern $P$ of length $m$ in $\LA$ in $O(m/\alpha + \log n)$ time where $\alpha$ is the number of characters we can pack in a word.
\end{lemma}

In the algorithm by Manber and Myers we maintain the left and right boundaries of the current search interval of $\LA$ denoted by $L$ and $R$ and the longest common prefix between $\LA[L]$ and $P$, and between $\LA[R]$ and $P$, that we denote by $l$ and $r$, respectively. Initially the search interval is the whole range of $\LA$ such that $L=1$ and $R=n$. In an iteration we do as follows: If $l=r$ we start comparing $\LA[M]$ with $P$ from index $l+1$ until we find a mismatch and update either $L$ and $l$, or $R$ and $r$, depending on whether $\LA[M]$ is lexicographic larger or smaller than $P$. 
Otherwise, when $l\not =r$, we perform an $\LCP$ query that enable us to either half the range of $\LA$ without reading from $P$ or start comparing $\LA[M]$ with $P$ from index $l+1$ as in the $l =r$ case. When $l>r$ there are three cases: If $\LCP(L, M)>l$ then $P$ is lexicographic larger than $\LA[M]$ and we set $L$ to $M$ and continue with the next iteration. If $LCP(L, M)<l$ then $P$ is lexicographic smaller than $\LA[M]$ and we set $R$ to $M$ and set $r$ to $\LCP(L,M)$ and continue with the next iteration. If $LCP(L, M)=l$ then we compare $\LA[M]$ and $P$ from index $l+1$ until we find a mismatch. Let that mismatch be at index $l+i$. If the mismatch means that $P$ is lexicographic smaller than $\LA[M]$ then we set $R$ to $M$ and set $r$ to $l+i-1$ and continue with the next iteration. If the mismatch means that $P$ is lexicographic larger than $\LA[M]$ then we set $L$ to $M$ and set $l$ to $l+i-1$ and continue with the next iteration. Three symmetrical cases exists when $r>l$.

\begin{figure}
\centering
\begin{tikzpicture}

\matrix (first) [table1withindices]
{
  \node[] (a) {$i+c'$}; & $\ldots$ & \node[] () {$i-1$}; &  \node[fill=black!20] (l1) {$i$}; & \node[fill=black!20] () {$\ldots$}; & \node[fill=black!20] (r1) {$i+c$};  & \node[text width=10pt] (b) {\texttt{000}}; & \node[fill=black!20] (c) {$i+c+1$}; & \node[fill=black!20] () {$\ldots$}; & \node[fill=black!20] (rr1) {$i+\alpha-1$}; & $i+\alpha$ & $\ldots$ & $i+c+\alpha$  & \node[text width=10pt] (d) {\texttt{000}}; \\
  \node[fill=black!20] (l2) {$i$};  &\node[fill=black!20] () {$\ldots$}; & \node[fill=black!20] (r2) {$i+c$}; & \node[] (s1l) {\texttt{000..000}}; & \node[] () {$\ldots$}; &  \node[] (s1r) {\texttt{000..000}};  & \node[text width=10pt] () {\texttt{000}}; & \node[] (s2l) {\texttt{000..000}}; & \node[] () {$\ldots$}; & \node[] (s2r) {\texttt{000..000}}; & \node[fill=black!20] (ll2) {$i+c+1$}; & \node[fill=black!20] () {$\ldots$};& \node[fill=black!20] (rr2) {$i+\alpha-1$}; & \node[ text width=10pt] (g) {\texttt{101}}; \\
};
\draw [decorate,decoration={brace,amplitude=10pt},xshift=0pt,yshift=0pt] ($(a.north west)+(0,3pt)$) -- ($(b.north east)+(0,3pt)$) node [black,midway,yshift=15pt] {\footnotesize $w_1$};
\draw [decorate,decoration={brace,amplitude=10pt},xshift=20pt,yshift=10pt] ($(c.north west)+(0,3pt)$) -- ($(d.north east)+(0,3pt)$) node [black,midway,yshift=15pt] {\footnotesize $w_2$};
\draw [dotted] (l1.south west) -- (l2.north west);
\draw [dotted] (r1.south east) -- (r2.north east);
\draw [dotted] (c.south west) -- (ll2.north west);
\draw [dotted] (rr1.south east) -- (rr2.north east);
\draw [decorate,decoration={brace,amplitude=10pt,mirror}] ($(s1l.south west)+(0,-3pt)$) -- ($(s1r.south east)+(0,-3pt)$) node [black,midway,yshift=-15pt] {\footnotesize $s_1$};
\draw [decorate,decoration={brace,amplitude=10pt,mirror}] ($(s2l.south west)+(0,-3pt)$) -- ($(s2r.south east)+(0,-3pt)$) node [black,midway,yshift=-15pt] {\footnotesize $s_2$};
\draw [decorate,decoration={brace,amplitude=5pt,mirror}] ($(g.south west)+(0,-3pt)$) -- ($(g.south east)+(0,-3pt)$) node [black,midway,yshift=-15pt] {\footnotesize $g$};
\end{tikzpicture}
\caption{\label{fig:align}Alignment of $\alpha$ characters that extends over a word boundary where $c'=c+1-\alpha$. The relevant part of the lower word $w_1$ and upper word $w_2$ is combined with bitwise shifts, a bitwise or and the $g$ bits on the right is set to \texttt{0}.}
\end{figure}
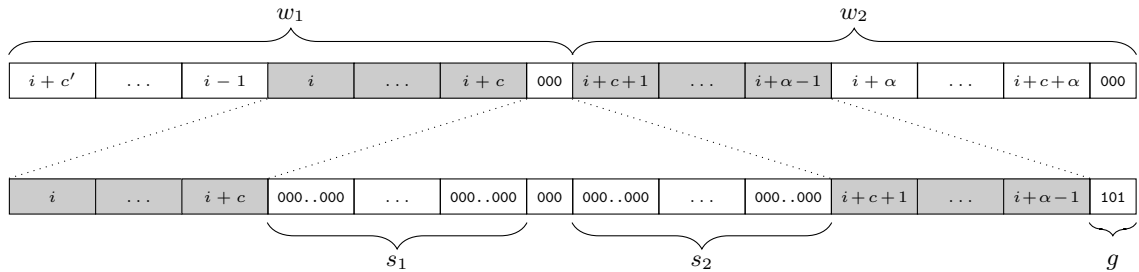

We generalize their algorithm to work on word packed strings such that we can compare $\alpha$ characters in constant time. In each iteration where we need to read from $P$ we align the next $\alpha$ characters from $P$ and $\LA[M]$ such that we can compare them in constant time: Assume that we need to read the range from $i$ to $i+\alpha-1$ in $P$. If this range of characters is contained in one word we do not need to align. Otherwise, 
we extract the relevant parts of the words that contain the range with bitwise shifts and combine them in $w_{align}$ with a bitwise or. See Figure \ref{fig:align}. We align the $\alpha$ characters from $\LA[M]$ in the same way and store them in $w'_{align}$.

We use a \emph{bitwise exclusive or} operation between $w_{align}$ and $w'_{align}$ to construct a word where the most significant set bit is at a bit position that belong to the mismatching character with the lowest index.
We obtain the position of the most significant set bit in constant time with the technique of Fredman and Willard~\cite{fredmanwillardfusion}. From this we know exactly how many of the next $\alpha$ characters that match and we can increase $i$ accordingly. Since every mismatch encountered result in a halving of the search range of $\LA$ we can never read more than $O(\log n)$ incomplete chunks. The number of complete chunks we read is bounded by $O(m/\alpha)$. Overall we obtain a $O(m/\alpha + \log n)$ time algorithm for matching in $\LA$. This result is summarized in Lemma \ref{lem:la}.

\subsection{Handling short patterns}\label{sec:tab}
Now we show how to answer count, locate and lexicographic predecessor queries on short patterns. We store an array containing an index for every possible pattern $P$ where $m< \log_\sigma (n)-1$ and at the index we store a pointer to the deepest node in $\T$ that prefix $P$. We call this node $d_P$.
We use $d_P$ as the basis for answering every query on short patterns.
We assume that the range in $\LA$ spanned by $d_P$ goes from $l$ to $r$. 
We answer predecessor queries as follows: If $P$ is lexicographic smaller than $\LA[0]$ then $P$ has no predecessor in $\LA$. Otherwise, we find the predecessor as follows: If $d_P$ is representing $P$ then the predecessor of $P$ is located at index $l-1$ of $\LA$. Otherwise, we assume that $d_P$ prefix $P$ with $i$ characters and need to decide whether $P$ continues on an edge out of $d_P$ or $P$ deviates from $\T$ in $d_P$. 
We do this by querying the predecessor data structure over the children of $d_P$ with character $i+1$ of $P$. If this query does not return an edge, then $P[i+1]$ is lexicographic smaller than the first character of every edge out of $d_P$, and the predecessor of $P$ is the string located at index $l-1$ of $\LA$. If this query returns an edge $e_{pred}$ then there are two cases. 

Case 1: The first character of $e_{pred}$ is not identical to $P[i+1]$. Then the predecessor of $P$ is the lexicographic largest string in the subtree under $e_{pred}$. 

Case 2: The first character on $e_{pred}$ is identical to $P[i+1]$. In this case, if there exists an edge $e'_{pred}$ out of $d_P$ on the left side of $e_{pred}$, then the predecessor of $P$ is the lexicographic largest string in the subtree under $e'_{pred}$ and otherwise the predecessor is the string at index $l-1$ of $\LA$. 
We report the node in $\T$ that represents the predecessor of $P$. 

We let $e_{pred}$ be defined as above and answer count queries as follows: If $d_P$ represents $P$ we return the number of leafs spanned by $d_P$ in $\T$. If $P$ instead continues and ends on $e_{pred}$ we report the number of leafs spanned by the subtree below $e_{pred}$. We answer locate queries in the same way but instead of reporting the range we report the strings in the range.

We find $d_P$ in $O(1)$ time and $e_{pred}$ in $O(\log \log \sigma)$ time. In total we answer predecessor and count queries in $O(\log \log \sigma)$ time and locate queries in $O(\log \log \sigma + \occ)$ time

Since $m < \log_\sigma(n)-1$ there exists $\sigma+\sigma^2+\ldots+\sigma^{\lfloor \log_\sigma (n)-1 \rfloor}\leq \sigma^{\lfloor \log_\sigma (n) \rfloor}\leq \sigma^{\log_\sigma n}=n$ short patterns and we compute them in $O(n)$ time by performing a preorder traversal of $\T$ bounded to depth $\log_\sigma(n)-1$. Let $d_P$ be the node we are currently visiting and let $d_{next}$ be the node we visit next.  When we visit $d_P$ we fill the tabulation array for every string that is lexicographic larger or equal to the string represented by $d_P$ and lexicographic smaller than the string represented by $d_{next}$. We fill each of these indices with a pointer to $d_P$ since $d_P$ is the deepest node in $\T$ that represents a string that prefix these strings. We can store the tabulation array in $O(n)$ space.

\subsection{Handling long patterns}
Now we show how to answer count, locate and lexicographic predecessor queries on long patterns. We first give an overview of our solution followed by a detailed description of the individual parts. In $\T$ we distinguish between \emph{light} and \emph{heavy} nodes. If a subtree under a node spans at least $\log^2\log n$ leafs, we call the node heavy, otherwise we call it light. A node is a heavy branching node if it has at least two heavy children and all the heavy nodes constitutes a subtree that we call the heavy tree. We decompose the heavy tree into micro trees of height $\alpha$ and we augment every micro tree with a data structure that enables navigation from root to leaf in constant time. For micro trees containing a heavy branching node we do this with deterministic hashing and for micro trees without a heavy branching node we just compare the relevant part of $P$ with the one unique path of the heavy tree that goes through the micro tree. To avoid navigating the light nodes we in each light node store a pointer to the range of $\LA$ that the node spans. We construct two predecessor data structures for each micro tree: The \emph{light predecessor} structure over the strings represented by the light nodes that are connected to the heavy nodes in the micro tree and the \emph{heavy predecessor} structure over the heavy nodes in the micro tree. We answer queries on $P$ as follows: We traverse the heavy tree in chunks of $\alpha$ characters until we are unable to traverse a complete micro tree. This means that $P$ either continues in a light node, ends in the micro tree or deviates from $\T$ in the micro tree. We can decide if $P$ continues in a light node with the light predecessor structure and if this is the case we answer the query with the packed matching algorithm on the range of $\LA$ spanned by the light node. Otherwise, we use the heavy predecessor structure for finding $d_P$ in the micro tree and use $d_P$ for answering the query as in section \ref{sec:tab}.
The following sections describes in more detail how we build our data structure and answer queries and gives a time and space .

\subsubsection{Data structure}
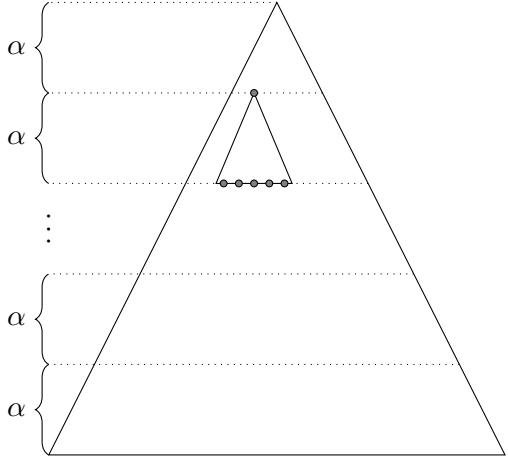
\begin{figure}
\centering
\begin{tikzpicture}[level 2/.style={sibling distance=25mm},level 1/.style={level distance=20mm}]
\tikzstyle{every node}=[circle]
\tikzset{
    subtree/.style={isosceles triangle, draw, inner sep=5pt,
anchor=apex, shape border rotate=90}
}

\coordinate (top) at (0,0) ;
\coordinate (lefttop) at (-3,0) ;
\coordinate (right) at (3,-6);
\coordinate (left) at (-3,-6) ;

\draw[] (top.south) -- (right) -- (left) -- cycle ;
\draw[dotted] ($ (top) !0.0! (right)$) -- ($ (lefttop) !0.0! (left)$) ;
\draw[dotted] ($ (top) !0.2! (right)$) -- ($ (lefttop) !0.2! (left)$) ;
\draw[dotted] ($ (top) !0.4! (right)$) -- ($ (lefttop) !0.4! (left)$) ;
\draw[dotted] ($ (top) !0.6! (right)$) -- ($ (lefttop) !0.6! (left)$) ;
\draw[dotted] ($ (top) !0.8! (right)$) -- ($ (lefttop) !0.8! (left)$);

\coordinate (basebase1) at ($($ (top) !0.2! (right)$) !0.5! ($ (top) !0.2! (left)$)$);
\coordinate (base1) at ($(basebase1) - (0.3,0)$);
\coordinate (alpha) at ($($($ (top) !0.2! (right)$) !0.5! ($ (top) !0.2! (left)$)$) - ($($ (top) !0.4! (right)$) !0.5! ($ (top) !0.4! (left)$)$)$);
\draw[] (base1) -- +($(-0.5,0) - (alpha)$) -- +($(0.5,0) - (alpha)$) -- cycle;
\node[draw,fill=gray, scale=0.25] at (base1) () {};
\node[draw,fill=gray, scale=0.25] at ($(base1) - (alpha)$) () {};
\node[draw,fill=gray, scale=0.25] at ($(base1) - (alpha) - (0.4, 0)$) () {};
\node[draw,fill=gray, scale=0.25] at ($(base1) - (alpha) - (0.2, 0)$) () {};
\node[draw,fill=gray, scale=0.25] at ($(base1) - (alpha) + (0.4, 0)$) () {};
\node[draw,fill=gray, scale=0.25] at ($(base1) - (alpha) + (0.2, 0)$) () {};


\draw [decorate,decoration={brace,amplitude=5pt},xshift=0pt,yshift=0pt]  (left) -- ( $(lefttop) !0.8! (left)$ ) node [midway,xshift=-12pt] () {$\alpha$};
\draw [decorate,decoration={brace,amplitude=5pt},xshift=0pt,yshift=0pt]  ( $(lefttop) !0.8! (left)$ ) -- ( $(lefttop) !0.6! (left)$ ) node [midway,xshift=-12pt] () {$\alpha$};
\draw [draw=white,decorate,decoration={brace,amplitude=5pt},xshift=0pt,yshift=0pt]  ( $(lefttop) !0.6! (left)$ ) -- ( $(lefttop) !0.4! (left)$ ) node [midway,xshift=0pt,rotate=-90] () {$\ldots$};
\draw [decorate,decoration={brace,amplitude=5pt},xshift=0pt,yshift=0pt]  ( $(lefttop) !0.4! (left)$ ) -- ( $(lefttop) !0.2! (left)$ ) node [midway,xshift=-12pt] () {$\alpha$};
\draw [decorate,decoration={brace,amplitude=5pt},xshift=0pt,yshift=0pt]  ( $(lefttop) !0.2! (left)$ ) -- ( $(lefttop) !0.0! (left)$ ) node [midway,xshift=-12pt] () {$\alpha$};


\end{tikzpicture}
\caption{The decomposition of $\HT$ in micro trees of height $\alpha$. One micro tree is shown with the root at string depth $\alpha$ and the boundary nodes at string depth $2\alpha$\label{fig:decomposition}}
\end{figure}

This section describes our data structure in details.
If a subtree under a node in $\T$ spans at least $\log^2\log n$ leafs, we call the node heavy. The heavy tree $\HT$ is the induced subgraph of all the the heavy nodes in $\T$. We decompose $\HT$ into \emph{micro trees} of string depth $\alpha$. A node, explicit or implicit, is a boundary node if its string depth is a multiple of $\alpha$. Except for the original root and leafs of $\HT$, each boundary node belongs in two micro trees i.e., a boundary node at depth $d \alpha$ is root in a micro tree that starts at string depth $d\alpha$ and is a leaf in a micro tree that starts at string depth $(d-1)\alpha$.
Figure \ref{fig:decomposition} shows the decomposition of $\HT$ into micro trees of string depth $\alpha$.

We augment every micro tree with information that enables us to navigate from root to leaf in constant time. To avoid using too much space we promote only some of the implicit boundary nodes to explicit nodes. We distinguish between three kinds of micro trees:
\begin{itemize}
\item  \textbf{Type 1.} At least one heavy branching node exists in the micro tree:
We promote the root and leafs to explicit nodes and use deterministic hashing to navigate the micro tree from root to leaf. Because the micro tree is of height $\alpha$, each of the strings represented by the leafs in the micro tree fits in a word and can be used as a key for hashing. We say that the root is a hashing node and the leafs are hashed nodes. We will postpone the analysis of time and space used by the micro trees that use hashing for navigation.
\item \textbf{Type 2.} No heavy branching node exists in the micro tree:
When the micro tree does not contain a heavy branching node, the micro tree is simply a path from root to leaf. Here we distinguish between two cases:
\begin{itemize}
\item \textbf{Type 2a.} The micro tree contains a non branching heavy node:

We promote the root and leaf to explicit nodes. Navigating from root to leaf takes constant time by comparing the string represented by the leaf with the appropriate part of $P$. We charge the space increase from the promotion of the root and leaf to the non branching heavy node. Since there are at most $n$ non branching heavy nodes we never promote more than $2n$ implicit nodes from type 2a micro trees.
\item \textbf{Type 2b.} The micro tree does not contain a heavy node:
  If the root is a boundary node where the micro tree above contains a heavy node we promote the root to an explicit node and store a pointer to the root of the nearest micro tree below that contains a heavy node. The path from root to root corresponds to a substring in $S$ and we navigate by comparing this string to the appropriate part of $P$. We charge the space increase from the promotion of the root to the heavy node descendant. Since we have at most $n$ heavy nodes we promote no more than $n$ implicit nodes from type 2b micro trees. We ignore every micro tree where the micro tree above does not contain a heavy node.
\end{itemize}
\end{itemize}

We say that a node in $\T$ is a heavy leaf if it is a heavy node with no heavy children. We want to bound the number of heavy branching nodes and heavy leafs. Every heavy leaf spans at least $\log^2\log n$ leafs of $\T$. This means we can have at most $n/\log^2\log n$ heavy leafs in $\T$. Since we have at most one branching heavy node per heavy leafs the number of heavy branching nodes is at most $n/\log^2\log n$.

We want to bound the number of implicit nodes that are promoted to explicit hashed nodes. This number is critical for constructing all hash functions in $O(n)$ time. We bound the number of promoted hashed nodes by associating each with the nearest descendant that is either a heavy branching node or a heavy leaf: Let $l$ be a promoted hashed node in a micro tree that contain a heavy branching node $h$. Then every promoted hashed node above $l$ is associated with $h$ or a node above $h$ in the tree. Hence, no other promoted node can be associated with the first encountered heavy branching or leaf node below $l$. Since we have at most $O(n/\log^2\log n)$ heavy branching and heavy leaf nodes we also have at most $O(n/\log^2\log n)$ implicit nodes that are promoted to explicit hashed nodes.

With deterministic hashing from Lemma \ref{lem:hash} the total time for constructing the explicit hashing nodes are:  
\begin{gather*}
O\left(\sum_{h\in H} |h|\log^2\log |h|\right) = O\left(\sum_{h\in H} |h|\log^2\log (n/\log^2\log n)\right)\\ = O\left(\log^2\log (n/\log^2\log n)\cdot \sum_{h\in H} |h|\right) = O\left(\log^2\log (n/\log^2\log n) \frac{n}{\log^2\log n}\right) = O(n)
\end{gather*}
Here $H$ is the set of all the hash functions and we bound the elements in every hash function $h$ to $n/\log^2\log n$.
Summing the elements of every hash function is bounded by the maximum number of promoted nodes, i.e. $O(n/\log^2\log n)$. To conclude, we spend linear time constructing the hash functions in the micro trees that contain a heavy branching node.

We associate two predecessor data structures with each micro tree that contains a heavy node: The first predecessor structure contains every light node that is a child of a heavy node in the micro tree. We call this predecessor data structure for the \emph{light predecessor structure} of the micro tree. The key for each light node is the string on the path from the root of the micro tree to the node itself padded with character $\$$ such that every string has length $\alpha$. These keys are ordered lexicographic in the predecessor data structure and a successful query yields a pointer to the node. The second predecessor structure is similar to the first but contains every heavy node in the micro tree. We call this predecessor structure for the \emph{heavy predecessor structure}.
We use Lemma \ref{lem:pred} for the predecessor structures. The total size of every light and heavy predecessor structures is $O(n)$ and a query in both take $O(\log \log n)$ because the universe is of size $(\sigma+1)^\alpha$.

For each light node that are a child of a heavy node we additionally store pointers to the range of $\LA$ that corresponds to the leafs in $\T$ that the light node spans.

\subsubsection{Answering queries}
We answer queries on long patterns as follows. First we search for the deepest micro tree in $\HT$ where the root prefix $P$. We do this by navigating the heavy tree in chunks of $\alpha$ characters starting from the root. Assuming that we have already matched a prefix of $P$ consisting of $i$ chunks of $\alpha$ characters we need to show how to match the $(i+1)$th chunk: If the micro tree is of type 1 and $P$ has length at least $(i+1)\alpha$, we try to hash the substring $P[i\alpha, (i+1)\alpha]$. If we obtain a node $v$ from the hash function we continue matching chunk $P[(i+1)\alpha, (i+2)\alpha]$ from $v$. If the micro tree is of type 2 we compare $\alpha$ sized chunks of $P$ with the string on the unique path from root to the first micro tree with an explicit root and continue matching from here. We have found the deepest micro tree where the root prefix $P$ when we are unable to match a complete chunk of $\alpha$ characters or are unable reach a micro tree with an explicit root. 
From this micro tree we need to decide whether the query is answered by searching $\LA$ from a light node or answered by finding $d_P$ in the micro tree, where $d_P$ is defined as in Section \ref{sec:tab}, i.e. the deepest node in $T_S$ that prefix $P$.
We check if $P$ continues in a light node by querying the light predecessor structure of the micro tree with the next unmatched $\alpha$ characters from $P$ and pad with character $\$$ if less than $\alpha$ characters remain unmatched in $P$. If the light node returned by the query represents a string that prefix $P$ we answer the query by searching the range of $\LA$ spanned by the light node with the packed matching algorithm. 

When $P$ does not continue in a light node we instead find and use $d_P$ for answering the query: If the micro tree is of type 2b or the root of the micro tree represents $P$ then $d_P$ is the root of the micro tree. Otherwise, we find $d_P$ by querying the heavy predecessor structure three times as follows: We call the remaining part of $P$, padded to length $\alpha$ with character $\$$, for $p_0$. We first query the predecessor structure with $p_0$ which yields a node that represents a string $n_0$. We then construct a string, $p_1$, that consists of the longest common prefix of $p_0$ and $n_0$, and as above, padded to length $\alpha$. We query the predecessor structure with $p_1$ which yield a new node that represents a string $n_1$. We then construct a string, $p_2$, that consists of the longest common prefix of $p_0$ and $n_1$, again padded to length $\alpha$. At last, we query the predecessor structure with $p_2$ which returns $d_P$. Given $d_P$, we answer count, locate and lexicographic predecessor queries exactly as we did in section \ref{sec:tab}.

Now we prove the correctness of our queries. First we prove that if $P$ continues in a light node then the query in the light predecessor structure returns that light node: Assume that $P$ goes through the light node $l_P$ that has a heavy parent in the micro tree $T_p$ and that we query the light predecessor structure with the string $Q_\alpha$. Let $L_{pred}$ be the string that represents $l_P$ in the light predecessor structure. Since $P$ goes through $l_P$ then $L_{pred}$ is identical or lexicographic smaller than $Q_\alpha$. Let $L'_{pred}$ be the successor of $L_{pred}$ in the light predecessor structure. Since $L_{pred}$ is lexicographic smaller than $L'_{pred}$ and has a longer common prefix with $Q_\alpha$ than $L'_{pred}$ has with $Q_\alpha$, then $L'_{pred}$ must be lexicographic larger than $Q_\alpha$. Since $Q_\alpha$ is identical or lexicographic larger than $L_{pred}$ and lexicographic smaller than $L'_{pred}$, a query on $Q_\alpha$ in the light predecessor structure will return $l_P$.

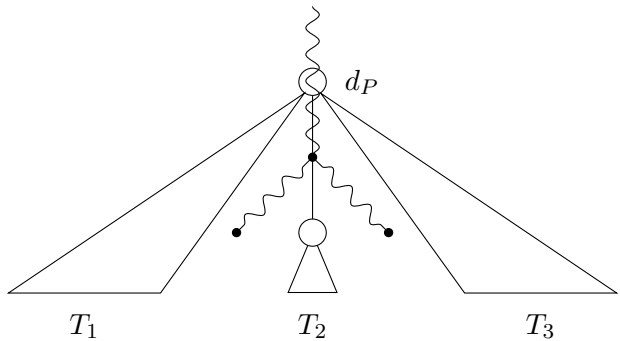
\begin{figure}
\centering
\begin{tikzpicture}[level 2/.style={sibling distance=25mm},level 1/.style={level distance=20mm}]
\tikzstyle{every node}=[circle]
\tikzset{
    subtree/.style={isosceles triangle, draw, inner sep=5pt,
anchor=apex, shape border rotate=90}
}

\node at (0,0) [solid,draw, label=right:$d_P$] (pv){}
     child [solid,child anchor=apex, label=below] {node [draw, subtree, label=below:$T_2$] (t3) {} node [draw,fill=white,label=right:$$] (v1) {}};
\draw[] (pv) -- ($ (t3.south) + (-2,0) $) -- node[below] {$T_1$} ++(-2,0) -- cycle ;
\draw[] (pv) -- ($ (t3.south) + (2,0) $) -- node[below] {$T_3$} ++(2,0) -- cycle ;
\draw[] decorate [decoration={name=snake}] {(0,1) -- (0,-1) node[fill=black,draw,scale=0.3] {}};
\draw[] decorate [decoration={name=snake,pre length=5pt}] {(0,-1) -- (-1,-2) node[fill=black,draw,scale=0.3] {}};
\draw[] decorate [decoration={name=snake,pre length=5pt}] {(0,-1) -- (1,-2) node[fill=black,draw,scale=0.3] {}};

\end{tikzpicture}
\caption{Searching for a prefix of $P$ in $\HT$\label{fig:deepest}}
\end{figure}

We now prove that the queries in the heavy predecessor structure always returns $d_P$: Because $P$ is not prefixed by a leaf of the micro tree or a light node from the light predecessor structure we know that $d_P$ is a heavy node in the micro trie. In Figure \ref{fig:deepest}, $d_P$ is depicted and $P$ either ends on or deviates from the edge $e$ that leads to the tree $T_2$. The trees $T_1$, $T_2$ and $T_3$ combined with $d_P$ and the edge $e$ constitutes the subtree of $d_P$. If $P$ deviates to the left or ends on $e$ then $P$ is lexicographic smaller than every string represented in $T_2$. If $P$ deviates to the right then $P$ is lexicographic larger than every string represented in $T_2$. Assume that $P$ deviates to the right on $e$. Then the query to the heavy predecessor structure with pattern $p_0$ will yield $n_0$ that represents the lexicographic largest string in $T_2$. The pattern $p_1$ will then be represented by the implicit node from where $P$ deviates from $e$. The pattern $p_1$ is lexicographic smaller than every string represented in $T_2$ and a query will yield $n_2$ as the lexicographic largest node in $T_1$ or, if $T_1$ is empty, the node $d_P$. Either way, the query on $p_2$ will yield the node $d_P$. We can make similar arguments for the other cases where $P$ ends on $e$, deviates left from $e$, ends at $d_P$ or goes through $d_P$ without following $e$.

The following gives an analysis of the running time of our queries. We spend at most $O(m/\alpha)$ time traversing the heavy tree. Both predecessor structures contains strings over a universe of size $n$ such that a query takes $O(\log\log n)$ time using Lemma \ref{lem:pred}. Each light node spans at most $\log^2\log n$ leafs which corresponds to an interval of length $\log^2\log n$ in $\LA$ that we search in $O(m/\alpha + \log \log \log n)$ time with the word accelerated algorithm for matching in $\LA$. Overall, we spend $O(m/\alpha + \log \log n)$ time for answering count and lexicographic predecessor queries and $O(m/\alpha + \log \log n + \occ)$ time for answering locate queries. 
Since we only query this data structure for patterns where $m\geq \log_\sigma(n)-1$ we have that $\log \log n = \log (\frac{\log n}{\log \sigma}\log\sigma) = \log\log_\sigma (n) + \log \log (\sigma) \leq \log (\log_\sigma (n) - 1) + 1 + \log \log (\sigma) \leq \log (m) + 1 + \log \log (\sigma)$, such that we answer count and lexicographic predecessor queries in $O(m/\alpha + \log m +\log\log \sigma)$ time and locate queries in $O(m/\alpha + \log m +\log\log \sigma + \occ)$ time. Combined with our solution for patterns where $m< \log_\sigma(n)-1$, that answer the queries in $O(\log \log \sigma)$ and $O(\log \log \sigma + \occ)$ time, respectively, we can for patterns of \emph{any} length answer count and lexicographic predecessor queries in $O(m/\alpha + \log m +\log\log \sigma)$ time and locate queries in $O(m/\alpha + \log m +\log\log \sigma + \occ)$ time. This is our main result which is summarized in Thm~\ref{theo:main}.

\bibliographystyle{splncs}
\bibliography{references}

\begin{thebibliography}{10}

\bibitem{Gusfield1997}
Gusfield, D.:
\newblock Algorithms on strings, trees, and sequences: computer science and
  computational biology.
\newblock Cambridge (1997)

\bibitem{manber1993suffix}
Manber, U., Myers, G.:
\newblock Suffix arrays: a new method for on-line string searches.
\newblock siam Journal on Computing \textbf{22}(5) (1993)  935--948

\bibitem{McCreight1976}
McCreight, E.M.:
\newblock A space-economical suffix tree construction algorithm.
\newblock J. ACM \textbf{23}(2) (1976)  262--272

\bibitem{Weiner1973}
Weiner, P.:
\newblock Linear pattern matching algorithms.
\newblock In: Proc. 14th Switching and Automata Theory. (1973)  1--11

\bibitem{FKS1984}
Fredman, M.L., Koml\'{o}s, J., Szemer{\'e}di, E.:
\newblock Storing a sparse table with 0(1) worst case access time.
\newblock J. ACM \textbf{31}(3) (1984)  538--544

\bibitem{FCFM2000}
Farach-Colton, M., Ferragina, P., Muthukrishnan, S.:
\newblock On the sorting-complexity of suffix tree construction.
\newblock J. ACM \textbf{47}(6) (2000)  987--1011

\bibitem{cole2006suffix}
Cole, R., Kopelowitz, T., Lewenstein, M.:
\newblock Suffix trays and suffix trists: structures for faster text indexing.
\newblock In: Automata, Languages and Programming.
\newblock Springer (2006)  358--369

\bibitem{fischer2015alphabet}
Fischer, J., Gawrychowski, P.:
\newblock Alphabet-dependent string searching with wexponential search trees.
\newblock In: Combinatorial Pattern Matching, Springer (2015)  160--171

\bibitem{Bille2011}
Bille, P.:
\newblock Fast searching in packed strings.
\newblock Journal of Discrete Algorithms \textbf{9}(1) (2011)  49--56

\bibitem{BBBGGW2014}
Ben-Kiki, O., Bille, P., Breslauer, D., Gasieniec, L., Grossi, R., Weimann, O.:
\newblock Towards optimal packed string matching.
\newblock Theoret. Comput. Sci. \textbf{525} (2014)  111--129

\bibitem{Belazzougui2012}
Belazzougui, D.:
\newblock Worst-case efficient single and multiple string matching on packed
  texts in the word-{RAM} model.
\newblock J. Disc. Algorithms \textbf{14} (2012)  91--106

\bibitem{arimura2016packed}
Takagi, T., Inenaga, S., Sadakane, K., Arimura, H.:
\newblock Packed compact tries: A fast and efficient data structure for online
  string processing.
\newblock In: Combinatorial Algorithms: 27th International Workshop, IWOCA
  2016, Helsinki, Finland, August 17-19, 2016, Proceedings. Volume 9843.,
  Springer (2016)  213

\bibitem{ruvzic2008constructing}
Ru{\v{z}}i{\'c}, M.:
\newblock Constructing efficient dictionaries in close to sorting time.
\newblock In: International Colloquium on Automata, Languages, and Programming,
  Springer (2008)  84--95

\bibitem{fredmanwillardfusion}
Fredman, M.L., Willard, D.E.:
\newblock Surpassing the information theoretic bound with fusion trees.
\newblock J. Comput. System Sci. \textbf{47}(3) (1993)  424--436

\end{thebibliography}

\end{document}